\documentclass[showpacs,preprintnumbers,amsmath,amssymb,floatfix]{revtex4-2}

\usepackage{graphicx}
\usepackage{dcolumn}
\usepackage{bm}
\usepackage{amssymb}
\usepackage{epsfig}
\usepackage{color}

\newcommand{\ba}{\begin{eqnarray}}
\newcommand{\ea}{\end{eqnarray}}
\newcommand{\be}{\begin{equation}}
\newcommand{\ee}{\end{equation}}
\newcommand{\bdisplay}{\begin{displaymath}}
\newcommand{\edisplay}{\end{displaymath}}

\newcommand{\eq}[1]{Eq.\,(\ref{#1})}
\newcommand{\fig}[1]{Fig.\,\ref{#1}}

\begin{document}

\title{Direct determination of the structure functions $F_L$, $F_S$ and $G$ from  $F_2$ and $dF_2/dQ^2$ to $O(\alpha_s^2)$}

\author{G.~R. Boroun}
\email{boroun@razi.ac.ir }
 \affiliation{Department of Physics, Razi University, Kermanshah
67149, Iran}

\author{ Loyal Durand}
\email{ldurand@hep.wisc.edu}
\altaffiliation{Present address: 415 Pearl Court, Aspen, CO 81611}
\affiliation{Department of Physics,
University of Wisconsin-Madison,
Madison, WI 53706}

\author{Phuoc Ha}
\email{pdha@towson.edu}
\affiliation{Department of Physics, Astronomy, and Geosciences, Towson University, Towson, MD 21252}

\begin{abstract}

We extend the results of Lappi {\em et al.}, Eur.~Phys.~J.~C {\bf 84}, 84 (2024), to show that it is possible to obtain expressions for the longitudinal,  singlet and gluon  structure functions $F_L$,  $F_S$ and $G$ in deep inelastic scattering directly in terms of the measured functions $F_2$ and $dF_2/\ln(Q^2)$ {\em modulo} non-singlet corrections expected to be small at very small $x$. The latter can be treated at low $x$ using existing quark distributions. Our results are presented consistently to $O(\alpha_s^2)$, correcting and extending the mixed-order results of Lappi {\em et al.}.

\end{abstract}


\maketitle

%%%%%%%%%%%%%%%%%%%%%%%%%
%%%%%%%%%%%%%%%%%%%%%%%%%

%%%%%%%%%%%%%%
%%%%%%%%%%%%%%

\section{Introduction \label{sec:Introduction}} 

%%%%%%%%%%%%%%

%\subsection{Derivation of $G$  and $F_S=x\Sigma$ from measured results for $F_2$ and $F_L$ \label{subsec:derivations}}

In a recent paper Lappi {\em et al.} \cite{Lappi}   showed that it is possible to determine the gluon and singlet structure functions $G=xg(x,Q^2)$ and $F_S(x,Q^2)=x\Sigma(x,Q^2)$ directly from the physical distributions $F_2(x,Q^2)$ and $F_L(x,Q^2)$ measured in deep inelastic lepton-proton scattering, and further derived evolution equations in $Q^2$ for $F_2$ and $F_L$ which involved only those physical quantities.  They demonstrated their method using an approximation in which they neglected the  non-singlet contributions $\Delta_i$ to the structure functions and retained only the first non-zero terms in the relevant QCD coefficient functions. Their results involve multiple convolutions and do not generalize simply in the form given.

Boroun and Ha \cite{BorounHa_FL} used the results in \cite{Lappi} to determine $F_L$, from the parametrization of the measured $F_2$ as a function of $x$ and $Q^2$  given by Block, Durand, and Ha (BDH) \cite{F2_parametrization}, with results which agreed reasonably well with existing measurements. However, the approximations used in \cite{Lappi} left open the question of how those results might be modified in a systematic expansion in powers of $\alpha_s$ which includes the complete  QCD coefficient functions and non-singlet effects, and how to develop such an expansion. 

Bl\"{u}mlein, Ravindran, and van Neervin  had earlier developed an expansion which can be used to determine  the singlet component of $F_L(x,Q^2)$ directly in terms of $F_2(x,Q^2)$ and $dF_2(x,Q^2)/d\ln{Q^2}$ \cite[Eqs.~36, 40-45]{Blumlein_1}, but did not carry their analysis to the point at which it could be used to predict $F_L$ from measured values of $F_2$ and its derivative. Their method could clearly also be applied separately to the singlet structure function $F_S(x,Q^2)$.

Kaptari, Kotikov, Chernikova, and Zhang \cite{Kaptari_1,Kaptari_2} later derived the equation for $F_L$ by a different method, and obtained results for that quantity correct to order $\alpha_s^2$ using the  complete coefficient functions and the parametrization for $F_2$ in \cite{F2_parametrization}. Their analysis using Mellin transforms \cite[\S~3]{Kaptari_2}  unfortunately involved approximations to the results in \cite{F2_parametrization} that, while probably sufficiently accurate, are difficult to assess.

Our objective here is to develop the method of Lappi {\em et al.} \cite{Lappi} systematically in a form in which one can use the complete coefficient functions calculated to any order in the strong coupling $\alpha_s(Q^2)$.   We will do this by using Laplace transforms to factor the convolutions which appear in the structure equations relating these functions. We then solve directly for the Laplace transforms of $G$ and $F_S$ in terms of the transforms of $F_2$, $F_L$, $\Delta$ and the coefficient functions in the structure equations.   The application of inverse Laplace transforms then gives the desired results without approximation as convolutions of known functions with $F_2$ and $dF_2/d\ln{Q^2}$ . As noted, Lappi {\em et al.} included only the first non-zero term in each of the coefficient functions; our analysis includes all terms to next-to-leading order in QCD. 

Our results can be implemented as in \cite{BorounHa_FL} using the very accurate fit to the HERA data on $F_2(x,Q^2)$ for $x\leq 0.1$ and $0.15 \leq Q^2\leq 3000$ GeV$^2$ as a function of both variables given by BDH \cite{F2_parametrization}.  The expression for $F_2(x,Q^2)$ used in that fit satisfies the analyticity requirements established by Ashok suri \cite{Suri}, connects smoothly to the cross section for real $\gamma p$ scattering, and respects the Froissart bound on the energy dependence of cross sections at high energies.  The results can be used to make predictions for experiments planned in regions with much smaller  $x$ and larger  $Q^2$ than have been explored so far.

%%%%%%%%%%%%%%%%%%%%%%%%%%

\section{Derivation of $G$  and $F_S=x\Sigma$ from measured results for $F_2$ and $F_L$ \label{sec:derivations}}

%%%%%%%%%%%%%%%%%%%%%%%%%%

We can  write the equations satisfied by the structure functions $F_2(x,Q^2)/x$ and $F_L(x,Q^2)/x$ in terms of the gluon distribution  $g(x,Q^2)$ and the singlet and non-singlet quarks distributions as
\ba
\label{sigma_defined}
\Sigma(x,Q^2) &=& \sum_i\left(q_{i,u}(x,Q^2)+q_{i,d}(x,Q^2)\right), \\
\label{delta_defined}
\Delta(x,Q^2) &=& \sum_i\left(q_{i,u}(x,Q^2)-q_{i,d}(x,Q^2)\right),
\ea
where the sums run over  $n_f$ generations of quarks and antiquarks.

In the presence of QCD corrections, $F_2$ and $F_L$ may be written as \cite{Curci,Furmanski,Ellisbook}
\ba
\label{tildeF2_eq}
F_2(x,Q^2)/x &=& \bar{e}_+^2\left(C_{F_2,\Sigma}\otimes \Sigma+C_{F_2,g}\otimes 2n_fg\right) +\bar{e}_-^2 C_{F_2,\Delta}\otimes\Delta \\
\label{tildeFL_eq}
F_L(x,Q^2)/x  &=&  \bar{e}_+^2\left(C_{L,\Sigma}\otimes\Sigma+C_{L,g}\otimes 2n_fg\right) + \bar{e}_-^2C_{F_L,\Delta}\otimes\Delta.
\ea
where the $C$s, calculated taking the quarks as effectively massless at high energies and small $x$,  are  $x$- and $Q^2$-dependent QCD coefficient functions, $\bar{e}_+^2$ and $\bar{e}_-^2$ are the average quark charges in the singlet (+) and non-singlet  (-) sectors, 
\be
\label{av_charges}
\bar{e}_+^2 =\sum_i\left(e_{i,u}^2+e_{i,d}^2\right)/2n_f,\quad \bar{e}_-^2 =\sum_i\left(e_{i,u}^2-e_{i,d}^2\right)/2n_f  
\ee
and $f\otimes g$ denotes the Mellin convolution \cite[\S 1.14(iv)]{dlmf}
\be
\label{otimes}
f\otimes g=\int_x^1 \frac{dz}{z} f(z)g\left(\frac{x}{z}\right).
\ee
We do not yet distinguish the orders of $\alpha_s$ in the $C$s which are sums of terms of increasing order.

While it can be argued  that it is a reasonable approximation to neglect the non-singlet contributions $\Delta$ to the structure functions at very small $x$ and large $Q^2$ where the quark distribution are dominated by sea quarks as was done by Lappi {\em et al.}, we will not do so as we may also wish to compare with  results at moderate values of the variables. We find in fact that the non-singlet contributions cancel to  first order in the strong coupling  $\alpha_s$ in our final expression for $F_L$ in terms of the measured structure function and its derivative $dF_2/d\ln{Q^2}$.

To proceed, we will first convert from the Mellin to  Laplace representations of the convolutions.  This is effected in Eqs.~(\ref{tildeF2_eq}) and (\ref{tildeFL_eq}) by the change of variables $x=e^{-v},\ z=e^{-w}$ with 
\be
f\otimes g=\int_0^v dw \widehat{f}(w)\widehat{g}(v-w),\quad \mathrm{where}\quad \widehat{h}(y)\equiv h(e^{-y}).
\ee
It will further be useful to change to a form of those expressions which involves the directly measurable functions $F_2$ and $F_L$, and the (derived) gluon and singlet quark distributions $G=xg$ and $F_S=x\Sigma=\sum_ix(q_i+\bar{q}_i)$ rather than $F_2/x$ and $F_L/x$. Since those expressions are of the form
\be
\frac{H(v)}{e^{-v}}  =  \sum_k\int_0^vdwC_k(w)\frac{F(v-w)}{e^{-(v-w)}},
\ee
simple multiplication by $e^{-v}$ and redefinition of the coefficient function with $C_k'(w)=e^{-w}C_k(w)$  gives the desired result, with $H(v)= \sum_kC_k'\otimes F$. 

With these redefinitions and with $x=e^{-v}$, Eqs.~(\ref{tildeF2_eq}) and (\ref{tildeFL_eq}) take the form
\ba
\label{F2_eq}
F_2(x,Q^2)  &=& \bar{e}_+^2\left(C_{F_2,F_S}\otimes F_S+C_{F_2,G}\otimes 2n_fG\right)+\bar{e}_-^2C_{F_2,\Delta'}\otimes\Delta'\\
\label{FL_eq}
F_L(x,Q^2)  &=& \bar{e}_+^2\left(C_{L,F_S}\otimes F_S+C_{L,G}\otimes 2n_fG\right)+\bar{e}_-^2C_{F_{L},\Delta'}\otimes\Delta', 
\ea
where 
\ba
C_{F_2,F_S}(w)&=&e^{-w}C_{F_2,\Sigma}(e^{-w}),\quad C_{F_2,G}(w)=e^{-w}C_{F_2,g}(e^{-w}), \\ 
C_{      L,F_S}(w)&=& e^{-w}C_{L,\Sigma}(e^{-w}), \quad C_{L,G}(w)=e^{-w}C_{L,g}(e^{-w}), \\ C_{F_2,\Delta'}(w)&=&e^{-w}C_{F_2,\Delta}(e^{-w}), \quad C_{F_L,\Delta'}(w)=e^{-w}C_{F_L,\Delta}(e^{-w}), \\  \Delta' &=&e^{-w}\Delta(e^{-w},Q^2). 
\ea
For notational simplicity, we do not indicate the $Q^2$ dependence of the coefficient functions which appears only through their dependence on $\alpha_s(Q^2)$.

Taking the Laplace transforms of the equations  factors the results through the relation (\cite{dlmf}, Sec.~1.14(iii)) 
\be
\mathcal{L}[f\otimes g;s]=\mathcal{L}[f;s]\mathcal{L}[g;s]
\ee
where
\be
\label{L_defined}
\mathcal{L}[h;s]\equiv \tilde{h}(s)=\int_0^\infty dy\,e^{-sy}h(y).
\ee
As indicated, we will use a tilde over a function symbol in the following to indicate the Laplace transform of that function.

The factored equations can be written compactly in matrix form as
\be
\label{matrix_form}
\mathsf{F}' =\bar{e}_+^2 \mathsf{C}\mathsf{V}  
\ee
where 
\be
\label{F',C_defined}
\mathsf{F}'=\begin{pmatrix}\tilde{F}_2'\\ \tilde{F}_L'\end{pmatrix} = \begin{pmatrix}
 \tilde{F}_2-\bar{e}_-^2\tilde{C}_{F_2,\Delta'}\tilde{\Delta}'  \\  \tilde{F}_L-\bar{e}_-^2\tilde{C}_{F_L,\Delta'} \tilde{\Delta'}
\end{pmatrix},\qquad
\mathsf{C}=\begin{pmatrix}
\tilde{C}_ {F_2,F_S} & \tilde{C}_{F_2,G}\\
 \tilde{C}_{L,F_S} &  \tilde{C}_{L,G}
\end{pmatrix},
\qquad \mathsf{V}=\begin{pmatrix}
\tilde{F}_S\\ 2n_f\tilde{G}
\end{pmatrix}.
\ee
The components of $\mathsf{F}'$ and $\mathsf{V}$ and  the coefficient functions in $\mathsf{C}$ are functions of  both the Laplace variable $s$ and of $Q^2$.

Inverting $\mathsf{C}$,
\be
\label{matrix_inverse}
\mathsf{V}=\mathsf{C}^{-1}\mathsf{F}'=\frac{1}{\bar{e}_+^2}\frac{1}{\mathrm{det}\mathsf{C}}\begin{pmatrix} \tilde{C}_{L,G} & - \tilde{C}_{F_2,G} \\
- \tilde{C}_{L,F_S} &  \tilde{C}_{F_2,F_S}
\end{pmatrix}\begin{pmatrix}
 \tilde{F}_2'\\ \tilde{F}_L'
\end{pmatrix},
\ee
or
\ba
\label{FS_result}
\tilde{F}_S(s,Q^2) &=&\frac{1}{\bar{e}_+^2}\frac{1}{\mathrm{det}\mathsf{C}}\left[ \tilde{C}_{L,G}(s)\tilde{F}'_2(s,Q^2)-\tilde{C}_{F_2,G}(s)\tilde{F}'_L(s,Q^2) \right] \\
\label{G_result}
\tilde{G}(s,Q^2)) &=& \frac{1}{2n_f\bar{e}_+^2}\frac{1}{\mathrm{det}\mathsf{C}}\left[\tilde{C}_{F_2,F_S}(s)\tilde{F}'_L(s,Q^2)-\tilde{C}_{L,F_S}(s)\tilde{F}'_2(s,Q^2)\right)].
\ea

These results are general and hold for coefficient functions $\tilde{C}_{a,b}$ given to any order in $\alpha_s$. 
The inverse Laplace transforms of these expressions give $F_S$ and $G$ through the general inverse relation (\cite{dlmf}, Sec.~1.14(iii),
\be
\label{Laplace_inverse}
h(t)=\mathcal{L}^{-1}[\tilde{h};t] = \frac{1}{2\pi i}\int_{-i\infty+c}^{i\infty+c}ds\, e^{st}\tilde{h}(s),
\ee
with $c$  chosen so that the integration contour lies to the right of the singularities of $\tilde{h}(s)=\mathcal{L}(h)(s)$ in the $s$ plane. 

We note that Lappi {\em et al.} did not use the factorization of the convolutions afforded by either the Laplace or Mellin transforms of the expressions in Eqs.~(\ref{tildeF2_eq}) and (\ref{tildeFL_eq}). As a result their initial solutions, {\em e.g.} for $dF_L/d\ln{Q^2}$, contain up to three convolutions. These cannot be reduced in general, but could be treated using partial integrations  in the special case of low-order polynomial coefficient functions accorded by their approximations.

Note that in the  usual approach to the QCD corrections, we would be working with Eqs. (\ref{F2_eq}) and (\ref{FL_eq}) which express  the physical (measurable)  quantities $F_2$ and  $F_L$ in terms of the derived quantities $F_S$, $G$, and $\Delta$.  The latter are obtained only indirectly in fits to the data with the QCD corrections  to a given order affecting those fits and the relation to $F_2$ and $F_L$. Here we are expressing $F_S$ and $G$ directly in terms of the measured $F_2$ and $F_L$ which can be taken as exact within the accuracy of the measured cross sections. The non-singlet correction $\Delta$ can be taken as negligible relative to $F_S$ at sufficiently small $x$ and large $Q^2$ where the singlet sea-quark distributions are expected to dominate.  Alternatively, to enable a comparison of our results with results at larger $x$ where $\Delta$ becomes important, we can estimate that function using quark distributions derived in other fits.

%%%%%%%%%%%%%%%%%%%%%%%%

\section{Expansion to $O(\alpha_s^2)$ \label{sec:o)alpha^2)}}

%%%%%%%%%%%%%%%%%%%%%%%%

To make use of the results above, we need to introduce the QCD expansions of the coefficient functions
\be
C_{a,b}(z)=\sum_k\left(\frac{\alpha_s}{2\pi}\right)^kC_{a,b}^{(k)}(z)
\ee
and their Laplace transforms,  and then expand the coefficient matrix  $\mathsf{C}^{-1}$ in \eq{matrix_inverse} to a specific order. The resulting expression gives the derived quantities $\tilde{F}_S$ and $\tilde{G}$ in terms of the measured functions $F_2$ and $F_L$  with QCD corrections to that order assuming that the non-singlet contributions are handled appropriately.

We will assume that the quarks can be treated as effectively massless with identical distributions when $Q^2$ is large and  $x$ is very small \cite{q_dist,BDHM2013}.  Then \cite{Curci,Furmanski,Ellisbook,ZvN3,Sanchez}, 
\ba
C_{F_2,F_S}(z) &=& \delta(1-z) + C_F\frac{\alpha_s}{2\pi}z\left[2\frac{\ln{(1-z)}}{(1-z)_+}-\frac{3}{2}\frac{1}{(1-z)_+}-(1+z)\ln{(1-z)} \right. \nonumber \\
\label{CF2FS}
&& \left. - \frac{1+z^2}{1-z}\ln{z}+3+2z-\left(\frac{9}{2}+\frac{\pi^2}{3}\right)\delta(1-z)\right] +\left(\frac{\alpha_s}{2\pi}\right)^2C_{F_2,F_S}^{(2)}(z)+\cdots,\\
\label{CF2G}
C_{F_2,G}(z) &=& \frac{1}{2}\frac{\alpha_s}{2\pi} z\left[\left(z^2+(1-z)^2\right)\ln{\left(\frac{1-z}{z}\right)}-8z^2+8z-1\right]+\left(\frac{\alpha_s}{2\pi}\right)^2C_{F_2,G}^{(2)}(z) + \cdots,\\
\label{CLFS}
C_{L,F_S}(z) &=& 2C_F\frac{\alpha_s}{2\pi}z^2+\left(\frac{\alpha_s}{2\pi}\right)^2C_{L,F_S}^{(2)}(z) +\cdots, \\
\label{CLG}
C_{L,G}(z) &=& 4T_R\frac{\alpha_s}{2\pi}z^2(1-z)+\left(\frac{\alpha_s}{2\pi}\right)^2C_{L,G}^{(2)}(z) +\cdots, \\
C_{F_2,\Delta'}(z) &=& \delta(1-z) + C_F\frac{\alpha_s}{2\pi}z\left[2\frac{\ln{(1-z)}}{(1-z)_+}-\frac{3}{2}\frac{1}{(1-z)_+}-(1+z)\ln{(1-z)} \right. \nonumber \\
\label{CF2Delta}
&& \left. - \frac{1+z^2}{1-z}\ln{z}+3+2z-\left(\frac{9}{2}+\frac{\pi^2}{3}\right)\delta(1-z)\right] +\left(\frac{\alpha_s}{2\pi}\right)^2C_{F_2,\Delta'}^{(2)}(z)+\cdots, \\
\label{CFLDelta}
C_{F_L,\Delta'}(z) &=&  2C_F\frac{\alpha_s}{2\pi}z^2+(\frac{\alpha_s}{2\pi})^2C_{F_L,\Delta'}^{(2)}(z)+\cdots,
\ea
where $z=e^{-w}$, $C_F=(N_c^2-1)/2N_c=4/3$ and $T_R=1/2$. The dependence of the coefficient functions on $Q^2$ through $\alpha_s$ is not indicated.  The terms of order $\alpha_s^2$ \cite{ZvN3,MochVermaseren} and $\alpha_s^3$ \cite{MVV,VMV}  are also known. 

The Laplace transforms of the coefficient functions are most simply calculated by changing  from $w$ to $z=e^{-w}$ as the integration variable, with 
\be
\label{simpleL}
\mathcal{L}(f)(s)=\int_0^1dz\,z^{s-1}f(z).
\ee
The distribution or generalized function $1/(1-z)_+$  in Eqs.~(\ref{CF2FS}) and (\ref{CF2G}) is defined as  \cite{Gelfand,LDPutikka}
\be
\label{+distribution}
\int_x^1 dz \frac{1}{(1-z)_+}f(z) \equiv \int_x^1 dz\frac{f(z)-f(1)}{1-z}+f(0)\ln(1-x).
\ee
Some results useful in the calculation of the transforms are given in the Appendix.

The Laplace transforms of the coefficient functions are
\ba
\tilde{C}_{F_2,F_S}(s) &=& 1- \frac{4}{3}\frac{\alpha_s}{2\pi}\left(\frac{9}{2}+\frac{\pi^2}{3}\right)+\frac{8}{3}\frac{\alpha_s}{2\pi} \left[ \left(\psi(1)-\psi(s+1)\right)^2+\psi'(1)-\psi'(s+1)  \right. \nonumber \\  
&& \left.-\left(\psi(1)-\psi(s+1)\right)
\left(\frac{3}{2}+\frac{1}{s+1}+\frac{1}{s+2}\right)+\frac{1}{(s+1)^2}+\frac{1}{(s+1)(s+2)}+\frac{1}{(s+2)^2}\right. \nonumber \\
\label{C_2S}
&& \left.+\frac{3}{s+1}+\frac{2}{s+3}\right]+ O(\alpha_s^2), \\
\tilde{C}_{F_2,G}(s) &=&  \frac{1}{2}\frac{\alpha_s}{2\pi}\left[\frac{1}{s+1}\left(\psi(1)-\psi(s+1)\right)-\frac{2}{s+2}\left(\psi(1)-\psi(s+2)\right)+\frac{2}{s+3}\left(\psi(1)-\psi(s+3)\right) \right.  \nonumber \\
&& \left. +\frac{1}{(s+1)^2}-\frac{2}{(s+2)^2}+\frac{2}{(s+3)^2}-\frac{1}{s+1}+\frac{8}{s+2}-\frac{8}{s+3}\right]+ O(\alpha_s^2)\nonumber\\
\label{C_2G}
&=& \frac{1}{2}\frac{\alpha_s}{2\pi}\left[\left(\frac{1}{s+1}-\frac{2}{s+2}+\frac{2}{s+3}\right)\left(\psi(1)-\psi(s+1)\right) +\frac{4}{s+2}-\frac{5}{s+3}\right]+ O(\alpha_s^2), \\
\label{C_LS}
\tilde{C}_{L,F_S}(s) &=& \frac{8}{3}\frac{\alpha_s}{2\pi}\frac{1}{s+2} + O(\alpha_s^2), \\
\label{C_LG}
\tilde{C}_{L,G}(s) &=& \frac{\alpha_s}{2\pi}\frac{2}{(s+2)(s+3)} + O(\alpha_s^2) \\
\tilde{C}_{F_2,\Delta'}(s) &=& 1- \frac{4}{3}\frac{\alpha_s}{2\pi}\left(\frac{9}{2}+\frac{\pi^2}{3}\right)+\frac{8}{3}\frac{\alpha_s}{2\pi} \left[ \left(\psi(1)-\psi(s+1)\right)^2+\psi'(1)-\psi'(s+1)  \right. \nonumber \\  
&& \left.-\left(\psi(1)-\psi(s+1)\right)
\left(\frac{3}{2}+\frac{1}{s+1}+\frac{1}{s+2}\right)+\frac{1}{(s+1)^2}+\frac{1}{(s+1)(s+2)}+\frac{1}{(s+2)^2}\right. \nonumber \\
\label{C_2Delta}
&& \left.+\frac{3}{s+1}+\frac{2}{s+3}\right]+ O(\alpha_s^2), \\
\label{C_LDelta}
\tilde{C}_{F_L,\Delta'}(s) &=& \frac{8}{3}\frac{\alpha_s}{2\pi}\frac{1}{s+2}+ O(\alpha_s^2) .
\ea

The determinant in \eq{matrix_inverse} can be expanded in powers of $\alpha_s$ as        
\ba
\label{det_expansion}
\mathrm{det}\mathsf{C} &=& \frac{\alpha_s}{2\pi}\tilde{C}_{L,G}^{(1)}+\left(\frac{\alpha_s}{2\pi}\right)^2\left(\tilde{C}_{L,G}^{(1)}\tilde{C}_{F_2,F_S}^{(1)}-\tilde{C}_{L,F_S}^{(1)}\tilde{C}_{F_2,G}^{(1)}+\tilde{C}_{L,G}^{(2)}\right)+O\left(\alpha_s^3\right).
\ea
Its inverse can be expanded accordingly with the leading term of order $\alpha_s^{-1}$ and the first correction of order $\alpha_s^0$,
\be
\label{alpha_s^0_result}
(\mathrm{det}\mathsf{C})^{-1} =\frac{2\pi}{\alpha_s}\left(\tilde{C}_{L,G}^{(1)} \right)^{-1}\left[1-\frac{\alpha_s}{2\pi}\tilde{C}_{F_2,F_S}^{(1)}+\frac{\alpha_s}{2\pi}\left(\tilde{C}_{L,F_S}^{(1)}\tilde{C}_{F_2,G}^{(1)}-\tilde{C}_{L,G}^{(2)}\right)\big/\tilde{C}_{L,G}^{(1)}+O\left(\alpha_s^2\right)\right].
\ee
Thus, if we suppose that the original Eqs.~(\ref{F2_eq}) and (\ref{FL_eq}) are correct to $O(\alpha_s)$ and retain this 
 accuracy in subsequent calculations, we find that 
 \be
 \label{order_alpha^0}
 \tilde{G}(s,Q^2) = \frac{2\pi}{\alpha_s}\frac{1}{2n_f\bar{e}_+^2}\left(\tilde{C}_{L,G}^{(1)} \right)^{-1}\left[\left(1+\frac{\alpha_s}{2\pi}\left(\tilde{C}_{L,F_S}^{(1)}\tilde{C}_{F_2,G}^{(1)}\right)/\tilde{C}_{L,G}^{(1)}\right)\tilde{F}'_L-\frac{\alpha_s}{2\pi}\tilde{C}_{L,F_S}^{(1)}\tilde{F}'_2\right].
 \ee

 In contrast, in illustrating their approach, Lappi {\em et al.} \cite{Lappi} retained only the leading non-zero term in each of the  the coefficient functions in Eqs.~(\ref{CF2FS}) to (\ref{CLG}) and inverted Eqs.~(\ref{tildeF2_eq}) and (\ref{tildeFL_eq})  to obtain an expression in which the $O(\alpha_s)$ contribution to the coefficient of $F'_L$ is missing. That construction led to the derivative expressions for $\Sigma$ and $g$ in their Eqs.~(22-25). Our expression instead follows from the usual convention of working to a fixed order in $\alpha_s$ in the calculation of the QCD corrections in the final results  given in  Eqs.~(\ref{F2_eq}) and (\ref{FL_eq}) or \eq{matrix_form}.
 
 More generally, taking the forms of Eqs.~(\ref{CF2FS})-(\ref{CLG}) into account, we can write the expression for $\tilde{G}(s,Q^2)$ in \eq{G_result} as
 \ba
 \label{G1}
 \tilde{G}(s,Q^2) &=& \frac{1}{\tilde{C}_{F_2,F_S}\tilde{C}_{L,G}-\tilde{C}_{L,F_S}\tilde{C}_{F_2,G}}\frac{1}{2n_f\bar{e}_+^2}\left(\tilde{C}_{F_2,F_S}\tilde{F}'_L-\tilde{C}_{L,F_S}\tilde{F}'_2\right)  \\
 \label{G2}
 &=& \frac{1}{\tilde{C}_{L,G}-\tilde{C}_{L,F_S}\tilde{C}_{F_2,G}/\tilde{C}_{F_2,F_S}}\frac{1}{2n_f\bar{e}_+^2}\left(\tilde{F}'_L-\frac{\tilde{C}_{L,F_S}}{\tilde{C}_{F_2,F_S}}\tilde{F}'_2\right) \\
 &=& \frac{2\pi}{\alpha_s}\frac{1}{2n_f\bar{e}_+^2}\frac{1}{\tilde{C}_{L,G}^{(1)}}\left[1+\frac{\alpha_s}{2\pi}\frac{1}{\tilde{C}_{L,G}^{(1)}}\left(\tilde{C}_{L,G}^{(2)}-\tilde{C}_{L,F_S}^{(1)}\tilde{C}_{F_2,G}^{(1)}\right)+O(\alpha_s^2)\right]\tilde{F}'_L \nonumber \\
  \label{G3}
 && +\frac{2\pi}{\alpha_s}\frac{1}{2n_f\bar{e}_+^2}\frac{1}{\tilde{C}_{L,G}^{(1)}}\left[\frac{\alpha_s}{2\pi}\tilde{C}_{L,F_S}^{(1)}+O(\alpha_s^2)\right]\tilde{F}'_2.
 \ea
 The inverse Laplace transform of this expression using the coefficient functions above gives $G$ as a function of $v=\ln{(1/x)}$. 
 
 We can solve similarly for $\tilde{F}_S(s,Q^2)$,  
 \ba
 \label{F_S1}
 \tilde{F}_S(s,Q^2) &=& \frac{1}{\tilde{C}_{F_2,F_S}\tilde{C}_{L,G}-\tilde{C}_{L,F_S}\tilde{C}_{F_2,G}}\frac{1}{e_+^2}\left(\tilde{C}_{L,G}\tilde{F}_2-\tilde{C}_{F_2,G}\tilde{F}'_L\right) \\
 \label{F_S2}
 &=& \frac{1}{\tilde{C}_{F_2,F_S}-\tilde{C}_{L,F_S}\tilde{C}_{F_2,G}/\tilde{C}_{L,G}}\frac{1}{e_+^2}\left(\tilde{F}'_2-\frac{\tilde{C}_{F_2,G}}{\tilde{C}_{L,G}}\tilde{F}'_L\right)  \\
 &=& \left\{1-\frac{\alpha_s}{2\pi}\left(\tilde{C}_{F_2,F_S}^{(1)}-\frac{1}{\tilde{C}_{L,G}^{(1)}}\tilde{C}_{L,F_S}^{(1)}\tilde{C}_{F_2,G}^{(1)}\right)-\left(\frac{\alpha_s}{2\pi}\right)^2\left[\tilde{C}_{F_2,F_S}^{(2)}-\frac{1}{\tilde{C}_{L,G}^{(1)}}\left(\tilde{C}_{L,F_S}^{(1)}\tilde{C}_{F_2,G}^{(2)}+\tilde{C}_{L,F_S}^{(2)}\tilde{C}_{F_2,G}^{(1)}\right)\right.\right. \nonumber \\
&& \left.\left.+\frac{1}{\left(\tilde{C}_{L,G}^{(1)}\right)^2}\tilde{C}_{L,F_S}^{(1)}\tilde{C}_{F_2,G}^{(1)}\tilde{C}_{L.G}^{(2)}-\left(\tilde{C}_{F_2,F_S}^{(1)}-\frac{1}{\tilde{C}_{L,G}^{(1)}}\tilde{C}_{L,F_S}^{(1)}\tilde{C}_{F_2,G}^{(1)}\right)^2\right] +O(\alpha_s^3) \right\} \nonumber \\
\label{F_S3}
&& \times\frac{1}{e_+^2}\left\{\tilde{F}'_2 -\left[\left(\tilde{C}_{F_2,G}^{(1)}+\frac{\alpha_s}{2\pi}\tilde{C}_{F_2,G}^{(2)}\right)\Big/\left(\tilde{C}_{L,G}^{(1)}+\frac{\alpha_s}{2\pi}\tilde{C}_{L,G}^{(2)}\right)\right]\tilde{F}'_L\right\}
 \ea
 where the coefficients of $\tilde{F}'_L$ can be expanded further to obtain a result valid to $O(\alpha_2^2)$.
 
 The results in Eqs.~(\ref{G3}) and (\ref{F_S3}) give the expansions of of the Laplace transforms of  $G$ and $F_S=x\Sigma$ in terms of the transforms of the observable quantities $F_2$ and $F_L$ and the non-singlet correction $\Delta$  when the QCD corrections in \eq{FS_result} and \eq{G_result} are known to $O(\alpha_s^2)$. The inverse Laplace transforms of these expressions give $G(e^{-v},Q^2)$ and $F_S(e^{-v},Q^2)$ with $x=e^{-v}$. 
 
 The structure of the corrections warrants some discussion.  Equations (\ref{G3}) and (\ref{F_S3}) express the functions $\tilde{F}_S$ and $\tilde{G}$  in terms of the Laplace transforms of the measurable quantities $F_2$ and $F_L$ in a series in $\alpha_s$.  The right-hand sides of those equations at a given order $n$ involve not only coefficient functions of  order $n$ but also products and ratios of coefficient functions of lower order with the combined orders summing to $n$, the result of expanding the formulas in Eqs.~(\ref{G1}) and (\ref{F_S1}). For example,  $\tilde{C}_{L,G}^{(2)}$ and $\tilde{C}_{L,F_S}^{(1)}\tilde{C}_{F_2,G}^{(1)}$ with total orders 2 appear  together as coefficients of $\alpha_s/2\pi$ in the first line of the expression for $G$, \eq{G3}. A number of similar and more complicated structures appear in \eq{F_S3}.  
 
 When the Eqs.~(\ref{G3}) and (\ref{F_S3}) are inverted to obtain $F_2$ and $F_L$, the combined structures disappear through $O(\alpha_s^2)$, the order of our expansion. The result is the original series for  $F_2$ and $F_L$  with only the coefficients $C_{i.j}^{(2)}$ appearing at  $O(\alpha_s^2)$. The first extra terms appear only at $O(\alpha_s^3)$, beyond the assumed accuracy of the original calculation.  The result generalizes at higher orders.

 %%%%%%%%%%%%%%%%%%%%%
 %%%%%%%%%%%%%%%%%%%%%
 
 \section{Determination of $F_L$ from $F_2$ and $dF_2/dQ^2$ \label{sec:FLdetermination}}
 
  %%%%%%%%%%%%%%%%%%%%%
  
 Lappi {\em et al.} \cite{Lappi} showed that it is possible to determine $F_L(x,Q^2)$ directly from the structure function $F_2(x,Q^2)$ and its derivative $dF_2(x,Q^2)/dQ^2$. However, the results in \cite{Lappi} were obtained using only the first nonzero term in each of the QCD coefficient functions, a serious restriction. In this section, we will sketch  the use of coefficient functions taken to a given order in $\alpha_s$  as above to relate $F_L$ to $F_2$ to that order. When combined with  the parametrization of all the existing HERA data on $F_2$ as a function of both $x$ and $Q^2$ given in \cite{F2_parametrization}, this allows the prediction of the behavior of $F_L$ as a function of $x$ and $Q^2$  prior to any extensive measurements of that quantity. 
 
 We begin with \eq{G_result} rewritten in the form
\be
\label{FL_eq2}
    \tilde{F}_L(s,Q^2) =\frac{1}{\tilde{C}_{F_2,F_S}(s)}\left[\tilde{C}_{L,F_S}(s)\tilde{F}_2'(s,Q^2)+2n_f\bar{e}_+^2\mathrm{det}\mathsf{C}\,\tilde{G}(s,Q^2))\right].
\ee
It was shown some time ago that $G(x,Q^2)$ can be determined to leading order directly from the evolution equation for $F_2(x,Q^2)$ either through an approach based on differential equations \cite{Diff_eq_G} or, more directly, using our Laplace transform method \cite{Analytic_solution_G}. We will follow the latter approach here. The result, used in \eq{FL_eq2}, will give an equation for $F_L$ in terms of $F_2$ and $dF/dQ^2$, both well determined in the HERA data.

The complete evolution equation for $F_2(x,Q^2)$ can be constructed to $O(\alpha_s^2)$ using the quark evolution equations given in \cite{Furmanski}, Sec.~5, and the results for the coefficient functions in \cite{Curci,Furmanski}. The results are complicated and involve two non-singlet contributions as well as the dominant singlet contribution at that order. We will not give the equations here but will simply illustrate our method using the leading-order evolution equation \cite{Gribov,Dokshitzer,AltarelliParisi}
\ba
\label{F_evolution1}
\frac{dF_2(x,Q^2)}{d\ln{Q^2}} &=& \frac{\alpha_s}{2\pi}\left[\bar{e}_+^2(F_S\otimes P_{qq}^{(1)}+2n_fG\otimes P_{qg}^{(1)} )+\bar{e}_-^2P_{qq,\Delta}^{(1)}\otimes\Delta\right]  \\
\label{F_evolution2}
&=& \frac{\alpha_s}{2\pi}\left[F_2\otimes P_{qq}^{(1)}+2n_f{\bar e}_+^2G\otimes P_{qg}^{(1)}\right]
\ea
where the $P_{ij}$ are the Altarelli-Parisi splitting functions \cite{AltarelliParisi} given for $C_F=4/3$ and $T_R=1/2$ by
\ba
\label{Pqq}
P_{qq}^{(1)} &=& \frac{4}{3}\left[\frac{3}{2}\delta(1-z)+\frac{1+z^2}{(1-z)_+}\right], \\
\label{Pqg} 
P_{qg}^{(1)}  &=& \frac{1}{2}\left[z^2+(1-z)^2\right],
\ea
with Laplace transforms
\ba
\label{Pqq(s)}
\tilde{P}_{qq}^{(1)}(s) &=& \frac{4}{3}\left(2\left(\psi(1)-\psi(s+1)\right)+\frac{3}{2}-\frac{1}{s+1}-\frac{1}{s+2}\right),  \\
\label{Pgq(s)}
\tilde{P}_{qg}^{(1)}(s) &=&  \frac{1}{2}\left(\frac{1}{s+1}-\frac{2}{s+2}+\frac{2}{s+3}\right).
\ea

We have noted in \eq{F_evolution2} that the singlet and non-singlet splitting functions in \eq{F_evolution1} are equal in leading order; they differ and a second non-singlet term enters at $O(\alpha_s^2)$.

After taking the Laplace transform of \eq{F_evolution2}, this becomes
\be
\label{LaplaceF2evol}
        \frac{d\tilde{F}_2(s,Q^2)}{d\ln{Q^2}} = \frac{\alpha_s}{2\pi}\left[\tilde{F}_2(s,Q^2)\tilde{P}_{qq}^{(1)}(s)+2n_f{\bar e}_+^2\tilde{G}(s,Q^2)\tilde{P}_{qg}^{(1)}(s)\right]
\ee
Solving for $\tilde{G}(s,Q^2)$, we find that \cite{Analytic_solution_G}
\be
\label{GfromF2}
\tilde{G}(s,Q^2)=\frac{1}{2n_f{\bar e}_+^2\tilde{P}_{qg}^{(1)}(s)}\left[\frac{2\pi}{\alpha_s}\frac{d\tilde{F}_2(s,Q^2)}{d\ln{Q^2}}-\tilde{P}_{qq}^{(1)}(s)\tilde{F}_2(s,Q^2)\right]
\ee

Using this result in \eq{FL_eq2} and expanding to $O(\alpha_s)$, we find that
\ba
\tilde{F}_L'(s,Q^2) &=& \frac{1}{\tilde{P}_{qg}^{(1)}}\left[\tilde{C}_{L,G}^{(1)}+\frac{\alpha_s}{2\pi}\left(-\tilde{C}_{F_2,G}^{(1)}\tilde{C}_{L,F_S}^{(1)}+\tilde{C}_{L,G}^{(2)}\right) \right]\frac{d\tilde{F}_2}{d\ln{Q^2}} \nonumber \\
\label{FL_from_F2_1}
&+& \frac{\alpha_s}{2\pi}\left(\tilde{C}_{L,F_S}^{(1)}\tilde{F}_2'-\tilde{C}_{L,G}^{(1)}\frac{\tilde{P}_{qq}^{(1)}}{\tilde{P}_{qg}^{(1)}}\tilde{F}_2\right)+\cdots,
\ea
where $\tilde{F}_2'$ and $\tilde{F}_L'$, defined in \eq{F',C_defined}, contain non-singlet contributions. We note, however, that the non-singlet terms in \eq{FL_from_F2_1} cancel between the two sides of the equation at $O(\alpha_s)$ because $\tilde{C}_{F_L,\Delta'}^{(1)}=\tilde{C}_{L,F_S}^{(1)}$. Thus, to first order,
\ba
\tilde{F}_L(s,Q^2) &=& \frac{1}{\tilde{P}_{qg}^{(1)}}\left[\tilde{C}_{L,G}^{(1)}+\frac{\alpha_s}{2\pi}\left(-\tilde{C}_{F_2,G}^{(1)}\tilde{C}_{L,F_S}^{(1)}+\tilde{C}_{L,G}^{(2)}\right) \right]\frac{d\tilde{F}_2}{d\ln{Q^2}} \nonumber \\
\label{FL_from_F2_2}
&+& \frac{\alpha_s}{2\pi}\left(\tilde{C}_{L,F_S}^{(1)}\tilde{F}_2-\tilde{C}_{L,G}^{(1)}\frac{\tilde{P}_{qq}^{(1)}}{\tilde{P}_{qg}^{(1)}}\tilde{F}_2\right).
\ea
The equality of coefficient functions fails at higher order, and non-singlet terms persist.

We emphasize that Eqs.~(\ref{GfromF2}) and (\ref{FL_from_F2_2}) involve only measurable physical quantities  {\em modulo} the presumably small non-singlet corrections at $O(\alpha_s^2)$ and higher, with the effects of the quark-gluon microstructure appearing only through the QCD coefficient functions.  No derived quark or gluon distributions appear as would be the case in standard expressions for the structure functions.  Thus a comprehensive fit to $F_2(x,Q^2)$ as a function of both variables as in \cite{F2_parametrization} leads to physical predictions for $F_L(x,Q^2)$, illustrating the program discussed by Lappi {\em et al.}.

%%%%%%%%%%%%%%%%%%%%%%%
%%%%%%%%%%%%%%%%%%%%%%%

\section{An application: $F_L$ to $O(\alpha_s)$ \label{sec_application}}

%%%%%%%%%%%%%%%%%%%%%%%

%%%%%%%%%%%%%%%%%

\subsection{Theoretical input   \label{subsec;input}}

%%%%%%%%%%%%%%%%%

In the following, we determine  $F_L(x,Q^2)$ correct to $O(\alpha_s)$ directly from the structure function $F_2(x,Q^2)$ and its derivative $dF_2(x,Q^2)/dQ^2$. We then compare our result for $F_L(x,Q^2)$ with the one presented in  \cite{BorounHa_FL} and with the existing data on $F_L$.  

 $\tilde{F}_L(s,Q^2)$ is given  correct to order $\alpha_s$  in \eq{FL_from_F2_1}, where we again note the cancellation and resultant absence of non-singlet contributions at this order. Ignoring the contribution from $\tilde{C}_{L,G}^{(2)}$  and using the expressions for the coefficient functions in Eqs.~(\ref{C_2S})-(\ref{C_LG}) we find that
\begin{eqnarray}
\label{FL_from_F2_3}
\tilde{F}_{L}(s,Q^2)&=&h_{1}(s)\frac{d\tilde{F}_{2}(s,Q^2)}{d{\ln}Q^2}+{h_{2}(s)}\tilde{F}_{2}(s,Q^2).
\end{eqnarray}
where the functions $h_{1}(s)$ and  $h_{2}(s)$ are given by
\begin{eqnarray} \label{hsi}
h_{1}(s)&=& 4 \left[\frac{1+s}{4+3s+ s^2}+\frac{2}{3} \frac{ \alpha_s(Q^2)}{2\pi}\left(\frac{-2-s+s^2}{(2+s)(4+3s+ s^2) }+\frac{\psi(s+1)-\psi(1)}{2+s}\right) \right], \nonumber\\
h_{2}(s)&=&\frac{16}{3} \frac{\alpha_s(Q^2)}{2\pi}\left(\frac{1-s+2(1+s) \left[\psi(s+1)-\psi(1)\right]}{4+3s+ s^2}\right).
\end{eqnarray}
Here $\psi(s)=\Gamma'(s)/\Gamma(s)$ is the digamma function,  
$\psi(1)=-\gamma$ where $\gamma$ is Euler's constant  $0.57721\cdots$. 

$F_L$ is given by the inverse Laplace transform of the expression in \eq{FL_from_F2_2}, 
\be
\label{FL(w)}
F_L(x,Q^2)=F_L(e^{-w},Q^2)\equiv\widehat{F}_L(w,Q^2)=\mathcal{L}^{-1}[\tilde{F}_L;w].
\ee
 Since the inverse transform of a product can be written as a convolution,
\be
\label{Linvers_FG}
\mathcal{L}^{-1}[\tilde{f}\,\tilde{g};w]=(f\otimes g)(w)=\int_0^wdvf(w-v)g(v),
\ee
we obtain $F_L$ as a convolution of the measured functions $F_2$ and $dF_2/d\ln{Q^2}$ with the inverse transforms of the coefficient functions,
\be
\label{FL_convolution}
\widehat{F}_L(w,Q^2)=\int_0^w dv\left[J_1(w-v,Q^2)\frac{dF_2(e^{-v},Q^2)}{d\ln{Q^2}}+J_2(w-v,Q^2)F_2(e^{-v},Q^2)\right],
\ee
with $J_i(v,Q^2)=\mathcal{L}^{-1}[h_i;v]$.

Alternatively,  writing the convolution in Mellin form,
\be
\label{FL_Mellin}
F_L(x,Q^2)=\int_x^1\frac{dz}{z}\left[J_1\left(\frac{x}{z}\right)\frac{dF_2(z,Q^2)}{d\ln{Q^2}}+J_2\left(\frac{x}{z}\right)F_2(z,Q^2)\right],
\ee

The calculation of $J_1$ and $J_2$ is straightforward. The
digamma functions in $h_1$ and $h_2$ have infinite strings of simple poles with residue $-1$ at negative integer vales of $s$, $s= -1,\,-2,\,\cdots$., with \cite[5.7.6]{dlmf}
\be
\label{psi_series}
\psi(s+1)-\psi(1) = -\frac{1}{s+1} + \sum_{k=1}^\infty\frac{s+1}{k(k+s+1)}.
\ee
 There are also simple poles in  $h_1$ and $h_2$ at the zeros of the denominator $(s^2+3s+4)$ at $s_\pm=\frac{1}{2}\left(-3\pm\ i\sqrt{7}\right)$, and a second-order pole in $h_1$ at $s=-2$ from the ratio $\psi(s+1)/(s+2)$. There are no other singularities in the left-half $s$ plane. The integrand for the inverse Laplace transform in \eq{Laplace_inverse} involves a further factor $e^{vs}$ so decreases exponentially for $\Re s\rightarrow -\infty$ for $v=\ln{(1/x)}>0$. We can therefore close the integration contours in the inverse transforms in the left-half $s$ plane and evaluate the integrals simply  as sums of the residues at the  poles. 
 
With the resulting kernels defined as
${J}_{i}(\upsilon){\equiv}{\mathcal{L}^{-1}}[h_{i}(s);\upsilon]$,
 we find that  
\begin{eqnarray}
J_{1}(\upsilon)&=&4\ e^{-\frac{3}{2}\upsilon}\left[\cos{\left(\frac{\sqrt{7}}{2}\upsilon\right)}-\frac{\sqrt{7}}{7}
\sin{\left(\frac{\sqrt{7}}{2}\upsilon\right)}\right]  \nonumber  \\
 &+ &  \frac{8}{3}\frac{ \alpha_s(Q^2)}{2\pi} \left(2e^{-2\upsilon}  -
 e^{-\frac{3}{2}\upsilon}\left[\cos{\left(\frac{\sqrt{7}}{2}\upsilon\right)}+\sqrt{7}
\sin{\left(\frac{\sqrt{7}}{2}\upsilon\right)}\right] +  f_1(\upsilon)\right)  ,
\end{eqnarray}
and
\begin{eqnarray}
J_{2}(\upsilon)&=&-\frac{16}{3}\frac{\alpha_{s}(Q^2)}{2\pi}e^{-\frac{3}{2}\upsilon}\bigg{[}\cos\left(\frac{\sqrt{7}}{2}\upsilon\right)
-5\frac{\sqrt{7}}{7}
\sin\left(\frac{\sqrt{7}}{2}\upsilon\right)\bigg{]}
+\frac{32}{3}\frac{\alpha_{s}(Q^2)}{2\pi} f_2(\upsilon),
\end{eqnarray}
where $f_1(v)$ and $f_2(v)$ are associated with the digamma functions,
\begin{eqnarray}
f_1(\upsilon) &=& {\mathcal{L}^{-1}}\left[\frac{\psi(s+1)-\psi(1)}{2+s},s;\upsilon\right] \nonumber\\
&=&  - e^{- \upsilon} + \left(1-\upsilon\right)e^{-2  \upsilon} + e^{-2  \upsilon}\sum_{m=1}^\infty \frac{e^{-m  \upsilon}}{m}  \nonumber \\
\label{f1_result} 
&=& - e^{- \upsilon} + \left(1-\upsilon\right)e^{-2  \upsilon} -e^{-2  \upsilon}\ln{(1-e^{-\upsilon})}, \\ \nonumber \\
\label{f2_result}
f_2(\upsilon) &=& {\mathcal{L}^{-1}}\left[\frac{(s+1)\left(\psi(s+1)-\psi(1)\right)}{4+3s+s^2},s;\upsilon\right]\nonumber\\
&=& \frac{s_++1}{s_*-s_-}\left(\psi(s_++1)-\psi(1)\right)e^{s_+\upsilon}+\frac{s_-+1}{s_--s_+}\left(\psi(s_-+1)-\psi(1)\right)e^{s_-\upsilon} \nonumber \\
&+& \sum_{k=1}^\infty\frac{k}{((k+1)^2-3(k+1)+4}e^{-(k+1)\upsilon}, \qquad s_\pm=\frac{1}{2}\left(-3\pm
i\sqrt{7}\right).
\end{eqnarray}

Since 
%,
\be
\label{num1}
\frac{s_\pm+1}{s_\pm-s_\mp}\left(\psi(s_\pm+1)-\psi(1)\right)=0.11819\pm1.31979i,
\ee
the first two terms in \eq{f2_result} are complex conjugates and
$f_2$ is real. With a change of summation variable $m=k+1$ \eq{f2_result}  reduces to
\be
\label{f2_2}
f_2(\upsilon) = 2e^{-\frac{3}{2}\upsilon}\left[0.11819 \cos{\frac{\sqrt{7}}{2}\upsilon}-1.31979\sin{\frac{\sqrt{7}}{2}\upsilon}\right] + \sum_{m=1}^\infty\frac{m-1}{m^2-3m+4}e^{-m\upsilon}.
\ee

The sum in this expression converges well for $\upsilon$ large or $x=e^{-\upsilon}$ small, but poorly for $\upsilon$ small. It is therefore useful for numerical calculations to subtract  (at least) the first two terms in the expansion of the coefficient of $e^{-m\upsilon}$ in powers of $1/m$ to obtain a more rapidly convergent series, and add then add the sums of  those terms back in analytically. This gives
\ba
f_2(v) &=& 2e^{-\frac{3}{2}\upsilon}\left[0.11819 \cos{\frac{\sqrt{7}}{2}\upsilon}-1.31979\sin{\frac{\sqrt{7}}{2}\upsilon}\right]  \nonumber \\
&-& \ln{\left(1-e^{-\upsilon}\right)}+2 Li_2\left(e^{-\upsilon}\right) +
\sum_{m=1}^\infty\left[\frac{2m-4}{m\left(m^2-3m+4\right)}-\frac{2}{m^2}\right]e^{-m\upsilon}
\ea
where $Li_2(z)$ is the dilogarithm function \cite[25.12.1]{dlmf}. The remaining series converges as $2/m^3$. The estimated remainder in the series after $M\gg 10$ terms  is $\approx 1/M^2$ for $\upsilon$ small.

The use of these results for $J_1(v)$ and $J_2(v)$ in \eq{FL_convolution} gives the longitudinal structure function $\widehat{F}_L$ as a function of $w=\ln({1/x)}$ and $Q^2$. The more familiar $x,\,Q^2$  form is given by \eq{FL_Mellin}, explicitly
\begin{eqnarray} 
\label{FL_res}
F_{L}(x,Q^2)&=&4 \int_{x}^{1}\frac{dz}{z} \frac{d{F}_{2}(z,Q^2)}{d{\ln}Q^2} \left \{ \left(\frac{x}{z}\right)^{3/2}\left[\cos{\left(\frac{\sqrt{7}}{2}{\ln}\frac{z}{x}\right)}-\frac{\sqrt{7}}{7}
\sin{\left(\frac{\sqrt{7}}{2}{\ln}\frac{z}{x}\right)}\right]\right. \nonumber\\
&-&  \frac{2}{3} \frac{ \alpha_s(Q^2)}{2\pi} \left(\frac{x}{z}  +
 \left(\frac{x}{z}\right)^{3/2}\left[\cos{\left(\frac{\sqrt{7}}{2}{\ln}\frac{z}{x}\right)}+\sqrt{7}
\sin{\left(\frac{\sqrt{7}}{2}{\ln}\frac{z}{x}\right)}\right] \right. \nonumber\\
&-& \left. \left. \left(\frac{x}{z}\right)^{2}   \bigg{[}3-{\ln}\frac{z}{x} -\ln(1-\frac{x}{z})\bigg{]} \right) \right \}  \nonumber\\
&-& \frac{16}{3} \frac{\alpha_{s}(Q^2)}{2\pi}\int_{x}^{1} \frac{dz}{z}F_{2}(z,Q^2) \left \{\left(\frac{x}{z}\right)^{3/2}\bigg{[}(0.52724\cos{\left(\frac{\sqrt{7}}{2}{\ln}\frac{z}{x}\right)} \right. \nonumber\\
&+&\left. 3.3893
\sin{\left(\frac{\sqrt{7}}{2}{\ln}\frac{z}{x}\right)} \bigg{]}-
2\sum_{m=1}^{\infty}\bigg{(}\frac{2(m-4)}{m(m^2-3m+4)}
-\frac{2}{m^2} \bigg{)}\left(\frac{x}{z}\right)^{m} \right. \nonumber \\
&-&\left. \bigg{(}2 Li_ 2 \left(\frac{x}{z}\right) -\ln(1-\frac{x}{z})\bigg{)} \right \}.
\end{eqnarray}
%

%%%%%%%%%%%%%%%%%

\subsection{Numerical results   \label{subsec;results}}

%%%%%%%%%%%%%%%%%

\begin{figure}[h]
\label{fig1new}
\includegraphics[width=0.7\textwidth]{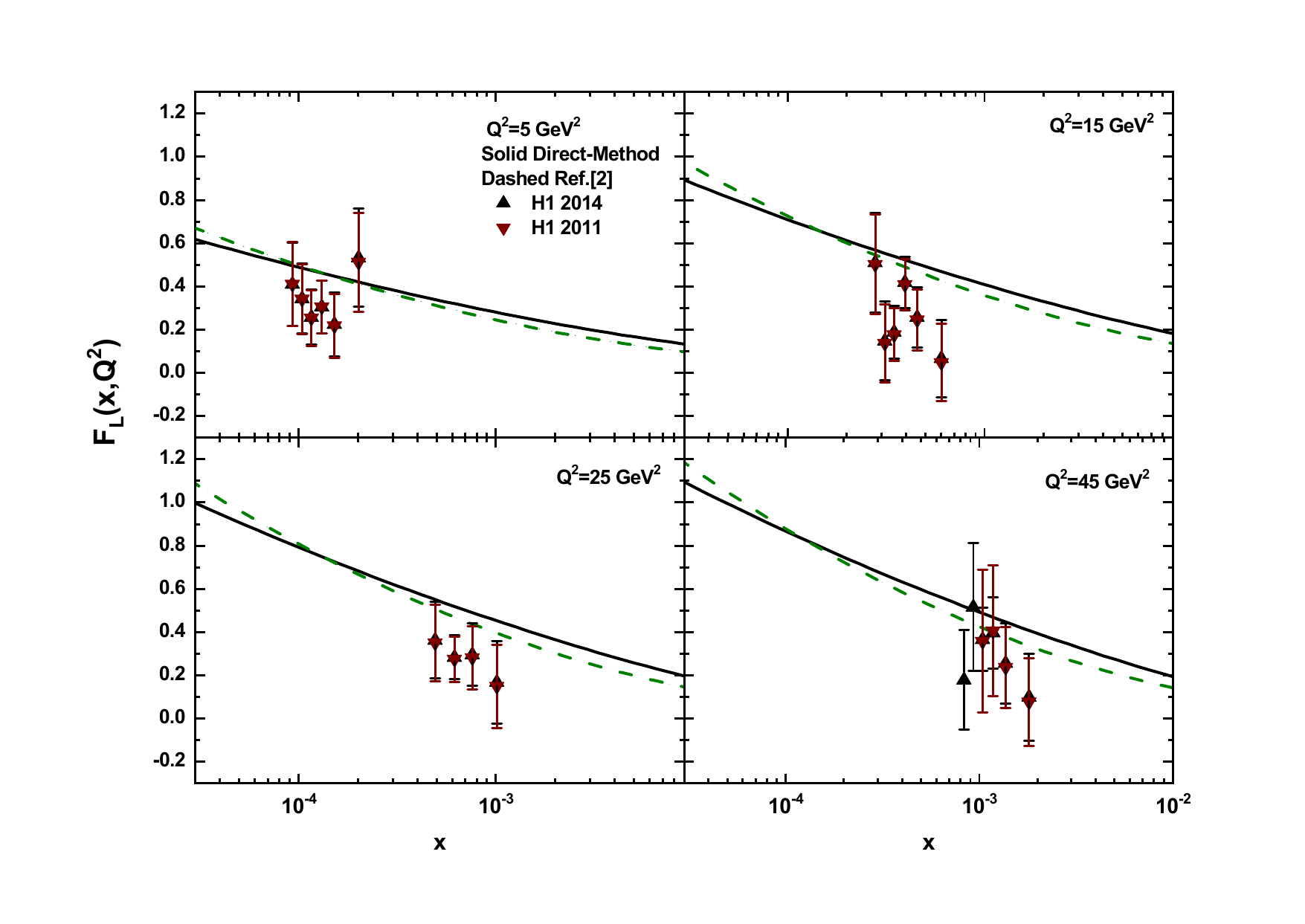}
\caption{The longitudinal structure function $F_{L}(x,Q^{2})$ calculated to $O(\alpha_s)$ from $F_2(x,Q^2)$ and $dF_2(x,Q^2)/d\ln{Q^2}$
as given by the comprehensive fit to the HERA data  of Block {\em et al.} \protect{\cite{F2_parametrization}}, plotted at fixed $Q^{2}$ as a function of $x$  (solid curves), compared to the results in \cite{BorounHa_FL} based on the approximations of Lappi {\em et al.} \protect{\cite{Lappi}} at LO (dashed
curves), and the H1 data from HERA \cite{H1_Coll_2011,H1_Coll_2014}. }\label{Fig1}
\end{figure}

To illustrate the direct derivation of $F_L$ from measured quantities, we have used the $O(\alpha_s)$ results in \eq{FL_res} to calculate $F_L(x,Q^2)$ from $F_2(x,Q^2)$ and $dF_2(x,Q^2)/d\ln{Q^2}$ as given by the very accurate comprehensive fit to the HERA data as a simultaneous function of $x$ and $Q^2$  obtained by Block, Durand, and Ha \cite{F2_parametrization}.
 In \fig{Fig1} we compare those results for $F_L(x,Q^2)$ (solid curves) with the  results obtained by Boroun and Ha  \cite{BorounHa_FL} using the approximate expressions of Lappi {\em et al.}  \cite{Lappi} (dashed curves) and with the H1 data from HERA at $Q^2 = 5, 15, 25$, and $45 \, {\rm GeV}^2$.  The differences in the complete and approximate leading order curves are small but significant.  The approximations of Lappi {\em et al.}, while essential in their approach, are not necessary when using the factorization of the corrections in Laplace space, and should not be made.
%
%%%%%%%%%%%%%%%%%%%%%%%
%
%
%%%%%%%%%%%%%%%%%%%%%%%

Kaptari {\em et al.} extended the calculation of $F_L$ from $F_2$ to include the QCD corrections of $O(\alpha_s^2)$ using their somewhat different method, and found that the inclusion of those higher-order corrections significantly improved the agreement with the data.  It would be of interest to extend the present calculations to that order  using the relations in Eqs.~(\ref{G3}) and (\ref{F_S3}) and  the similarly corrected evolution equation for $F_2$, thus   avoiding the approximations Kaptari {\em et al.} were forced to make in their approach. We note also that their approach did not involve a strict expansion of all contributions in powers of $\alpha_s$ so their $O(\alpha_s^2)$ calculation includes some probably-small higher order terms as well.

%%%%%%%%%%%%%%%%%%%%%%%
%%%%%%%%%%%%%%%%%%%%%%%

\section{Summary \label{sec:summary}}

 %%%%%%%%%%%%%%%%%%%%%%%
 
 We have presented  a systematic development  of the approach proposed by Lappi {\em et al.} \cite{Lappi} for evaluating the singlet structure function $F_S(x,Q^2)$ and the gluon structure function $G(x,Q^2)$ directly in terms of the measurable structure functions $F_2(x,Q^2)$ and $F_L(x,Q^2)$ in deep inelastic $ep$ scattering. The basic relations between these quantities are given Eqs.~(\ref{F2_eq}) and (\ref{FL_eq}). These express $F_2$ and $ F_L$ in terms of convolutions of $F_S$ and $F_L$ with a set of QCD corrections. Those equations were factored using Laplace transforms, eliminating the convolutions which appear in the original expressions, and could then be solved directly for the transforms of $F_S$ and $G$, Eqs.~(\ref{matrix_inverse})-(\ref{G_result}).
 
We assumed that the original equations Eqs.~(\ref{F2_eq}) and (\ref{FL_eq}) were correct to $O(\alpha_s^2)$, and expanded the QCD corrections in the solutions to that order. The results involved products and ratios of $O(\alpha_s)$ terms of combined order $\alpha_s^2$ which are necessary to ensure that the solutions reproduce the original equations to $O(\alpha_s^2)$ when transformed back to that form. The results for the Laplace transforms of $G$  and $F_S$ are given in Eqs.~(\ref{G3}) and (\ref{F_S3}) in terms of the Laplace transforms of $F_2$ and $F_L$ and of the QCD coefficient functions $C_{i,j}^{(k)}$, Eqs.~(\ref{C_2S})-(\ref{C_LDelta}). 

An inverse Laplace transform gives  $G$  and $F_S$ in terms of convolutions of  the measurable structure functions $F_2$ and $F_L$ with the inverse transforms of their coefficients. The results  here avoid the approximations in the coefficients made by Lappi {\em et al.} in their leading-order analysis and extend those results to $O(\alpha_s^2)$, and include the effects of non-singlet corrections. Unfortunately, the extant measurements of the longitudinal structure function $F_L$ are not extensive or accurate enough to make this approach practical even if the non-singlet terms are neglected.

As observed by Kaptari {\em et al.} \cite{Kaptari_1,Kaptari_2} and Lappi {\em et al.} \cite{Lappi}, the gluon distribution function $G$ in  \eq{G_result} can be expressed in terms of $F_2$ and $dF_2/d\ln{Q^2}$ using the evolution equation for $F_2(x,Q^2)$, a result demonstrated  some time ago in \cite{Diff_eq_G,Analytic_solution_G}.  The result, given to $O(\alpha_s)$ in  \eq{FL_from_F2_1}, gives the Laplace transform of $F_L$ directly in terms of the transforms of $F_2$ and $dF_2/d\ln{Q^2}$ and known QCD coefficients.  An  inverse Laplace transform then gives $F_L$ as a sum of convolutions of  the measurable $F_2$ and $dF_2/d\ln{Q^2}$ with the inverse transforms of the coefficients. An alternative approach was given in Bl\"{u}mlein {\em et al.} \cite{Blumlein_1}, but not carried to the point of application. 

This procedure was demonstrated at $O(\alpha_s)$ in Sec.~V. The inverse Laplace transforms of the coefficient functions were calculated analytically, and the final convolutions with $F_2$ and $dF_2/d\ln{Q^2}$ were calculated numerically using the very accurate parametrization of $F_2(x,Q^2)$ as a function of both variables given by Block, Durand, and Ha \cite{F2_parametrization}. The results are compared with existing data in Fig.~1. As noted above, this calculation has been extended to $O(\alpha_s^2)$ by Kaptari {\em et al.} using a different method that required  approximations in their final numerical calculations and eschewed an expansion in powers of $\alpha_s$, with results that improved the agreement with the data on $F_L$. It would clearly be of interest to extend the present calculation to that order, avoiding unnecessary approximations and including the non-singlet contributions which appear at that order.
  
  %%%%%%%%%%%%%%%%%%%%%%%
  
  \acknowledgements
  
  The authors would like to thank Dr.~J.~Bl\"{u}mlein for pointing out Ref.~\cite{Blumlein_1}.
  
  %%%%%%%%%%%%%%%%%%%%%%%

\appendix

 %%%%%%%%%%%%%%%%%%%%%%%
 
 \section{Some useful formulas \label{sec:formulas}}

We collect here some formulas useful in the calculation of the Laplace transforms required above. 
In particular, several nontrivial integrals can be expressed in terms of derivatives of the beta function 
$B(a,b)$  \cite[\S 5.12]{dlmf}, 
\be
\label{B(a,b)}
B(a,b) =  \int_0^1 dt\,t^{a-1} (1-t)^{b-1}=\frac{\Gamma(a)\Gamma(b)}{\Gamma(a+b)}.
\ee
Thus
\ba
%\int_0^1dt\,t^r\ln{t}&=&-\frac{1}{(r+1)^2}, \\
\int_0^1dt\,t^r\ln{(1-t)} &=& \frac{d}{dq}\int_0^1dt\,t^r(1-t)^q\Big\rvert_{q=0} = \frac{d}{dq}B(r+1,q+1)\Big\rvert_{q=0} \nonumber \\
&=& \frac{1}{r+1}\left(\psi(1)-\psi(r+2)\right),
\ea
where $\psi(r)$ is the digamma function, $\psi(r)=\Gamma'(r)/\Gamma(r)$ \cite[\S 5.1, \S 5.5]{dlmf} and $\psi(1)=-\gamma$ where $\gamma$ is Euler's constant $0.57721\cdots$. $\psi(r)$ satisfies the recurrence relation
\be 
\psi(r+n) = \frac{1}{r+n-1}+\cdots+\frac{1}{r}+\psi(r) 
\ee
which was used in the simplification of some results above. This integral can also be evaluated by expanding $\ln(1-t)$ as $\sum_{n=1}^\infty t^n/n$,  integrating term-by-term, and identifying the result with the  expansion of $\psi(z)$ in \cite[5.7.6]{dlmf}, using the recurrence relation.

A similar calculation gives
\ba
\int_0^1dt\,t^r\frac{\ln{t}}{1-t} &=& \int_0^1dt\left[rt^{r-1}\ln{t}\ln{(1-t)}+t^{r-1}\ln{(1-t)}\right] \nonumber \\
&=& \frac{d}{dq}\frac{d}{dr}\int_0^1dt\,rt^{r-1}(1-t)^q\Big\rvert_{q=0} \nonumber \\
&=&  \frac{d}{dq}\frac{d}{dr}\frac{\Gamma(r+1)\Gamma(q+1)}{\Gamma(r+q+1)} \Big\rvert_{q=0} \nonumber \\
&=& -\psi'(r+1),
\ea
a result that can also be extracted from \cite[4.251(4)]{Gradshteyn},

The same method can be used to evaluate ``+'' integrals with similar forms, with
\ba
\int_0^1dt\frac{t^r}{(1-t)_+} &=& \int_0^1dt\frac{t^r-1}{1-t} = r\int_0^1dt\,t^{r-1}\ln{(1-t)} \nonumber \\
&=& \psi(1)-\psi(r+1),
\ea
while
\ba
\int_0^1dt\,t^r\frac{\ln{(1-t)}}{(1-t)_+} &=& \int_0^1dt\frac{\ln(1-t)}{1-t}\left(t^r-1\right) \nonumber \\
&=& \frac{1}{2}r\int_0^1dt\,t^{r-1}\ln^2(1-t) 
= \frac{1}{2}r\frac{d^2}{dq^2}\int_0^1dt\,t^{r-1}(1-t)^q\Big\rvert_{q=0} \nonumber \\
&=& \frac{1}{2}\frac{d^2}{dq^2}\frac{\Gamma(r+1)\Gamma(q+1)}{\Gamma(r+q+1)}\Big\rvert_{q=0} \nonumber \\
&=& \frac{1}{2}\left[\left(\psi(1)-\psi(r+1)\right)^2+\psi'(1)-\psi'(r+1)\right]
\ea
where $\psi'(1)=\zeta(2)=\pi^2/6$. See also \cite[4.261(17)]{Gradshteyn}.

%\mathsf{G}(s,Q^2)) &=& \frac{1}{\mathrm{det}\mathsf{C}}\left[\mathsf{C}_{F_2,F_S}(s)\mathsf{F}_L(s,Q^2)-\mathsf{C}_{L,F_S}(s)\mathsf{F}_2(s,Q^2)\right)].
%

%%%%%%%%%%%%%%%%%%%%%%%%
 %%%%%%%%%%%%%%%%%%%%%%%


\begin{thebibliography}{99}

\bibitem{Lappi} T.~Lappi, H.~M\"{a}ntysaari, H.~Paukkunen, and M.~Tevio, ``Evolution of structure functions in momentum space," Eur.~Phys.~J.~C {\bf 84}, 84 (2024).

\bibitem{BorounHa_FL} G.~R.~Boroun and Phuoc Ha, ``Decoupling of the structure functions in momentum space based on the Laplace transform,'' Phys.~Rev.~D {\bf 109}, 094037 (2024).

\bibitem{F2_parametrization} M.~M.~Block, L.~Durand, and P.~Ha, ``Connection of the virtual $\gamma^*p$ cross section of $ep$ deep inelastic scattering to real $\gamma p$ scattering and the implications for $\nu N$ and $ep$ total cross sections," Phys.~Rev.~D {\bf 89}, 094027 (2014). The fit is to 395 datum points with $0.15\leq Q^2\leq3000$ GeV$^2$ and hadronic energies $W=\sqrt{Q^2(1-x)/x}>25$ GeV. The fit is excellent, with a $\chi^2$ per degree of freeedon of 0.95.

\bibitem{Blumlein_1} J.~Bl\"{u}mlein, V.~ Ravindran, and W.~L.~van Neerven, "On the Drell-Levy-Yan relation to $O(\alpha_s^2)$,  Nucl.~Phys.~B {\bf 586}, 349 (2000); arXiv:0004172 [hep-ph],  doi = 10.1016/S0550-3213(00)00422-3.

\bibitem{Kaptari_1} L.~P.~Kaptari, A.~V.~Kotikov, N.~Yu.~Chernikova, and Pengming~Zhang, ``Longitudinal Structure Function $F_L$ at small $x$
extracted from the Berger--Block--Tan Parametrization of $F_2$,''  JETP Lett. {\bf 109}, 281 (2019).

\bibitem {Kaptari_2} L.~P.~Kaptari, A.~V.~Kotikov, N.~Yu.~Chernikova, and Pengming~Zhang, ``Extracting the longitudinal structure function $F_L(x,Q^2)$ at small $x$ from a Froissart-bounded parametrization of $F_2(x,Q^2)$,''    Phys.~Rev.{\bf D} 99, 096019  (2019), https://doi.org/10.1103/PhysRevD.99.096019.

\bibitem{Suri} A.~Suri, Phys.~Rev.~D {\bf 4}, 570 (1971).

\bibitem{Curci}
G. Curci,  W.~Furmanski and R.~Petronzio, ``Lepton-hadron processes beyond leading order in quantum chromodynamics. The non-singlet case" Nucl~Phys.~B {\bf 175}, 27 (1980).

\bibitem{Furmanski} W.~Furmanski and R.~Petronzio, ``Lepton-hadron processes beyond leading order in quantum chromodynamics," Zeit.~Phys.~C {\bf 11}, 293 (1982).

\bibitem{Ellisbook} R.~K.~Ellis, W.~J.~Stirling, and B.~R.~Webber, ``QCD and Collider Physics,'' Cambridge University Press, Cambridge, England, 2003.

\bibitem{dlmf} {\it NIST Digital Library of Mathematical Functions},  F.~W.~J.~Olver, A.~B.~{Olde Daalhuis}, D.~W.~Lozier,  B.~I.~Schneider,  R.~F.~Boisvert, C.~W.~Clark, B.~R.~Miller, B.~V.~Saunders, H.~S.~Cohl, and M.~A.~McClain, eds., Release 1.2.6 of 2026-03-15 (2024).

%\bibitem{MellinLaplace} While the convolutions in \eq{otimes} can also be also factored using  Mellin transforms as is typically done, the use of Laplace transforms allows us to use the powerful and familiar techniques of complex integration in the evaluation our final expressions either analytically or numerically. The existence of extensive compilations of Laplace transforms is also advantageous.

\bibitem{q_dist} This not  true at accessible values of $x$ where the quark distributions continue to diverge from each other as $x$ decreases. This may be seen in \cite[Figs.\,2\ and 3]{BDHM2013}. However the fractional differences in the distributions do decrease, so the assumption of identical distributions becomes more accurate with decreasing $x$ and can be further improved through the use of  an effective value of the quark number $n_f$. 
\bibitem{BDHM2013} M.~M.~Block, L.~Durand, P.~Ha, and D.~W.~McKay, ``Implications of a Froissart bound saturation of $\gamma^*-p$ deep inelastic scattering. I. Quark distributions at ultra small $x$,"  Phys.~Rev.~D {\bf 88}, 014006 (2013).

\bibitem{ZvN3} E.~B.~Zijlstra and W.~L.~van Neerven, ``Deep inelastic QCD corrections to the deep inelastic structure functions $F_2$ and $F_L$," Nucl.~Phys.~B {\bf 383}, 525 (1992). 

\bibitem{Sanchez} J.~Sanchez Guill\'{e}n, J.~L.~Miramontes, M.~Miramontes, G.~Parente and O.~A.~Sampayo, ``Next-to-leading order analysis of the deep inelastic $R=\sigma_L/\sigma_T$", Nucl.~Phys.~B {\bf 353}, 337 (1991).

\bibitem{MochVermaseren} S.~Moch and J~.A~.M.~Vermaseren, ``Deep-inelastic structure functions at two loops," Nucl.~Phys.~B {\bf 573}, 853 (2000).

\bibitem{MVV} S.~Moch, J.~A.~M.~Vermaseren, and A.~Vogt, ``The longitudinal structure function at third order,'' Phys.~Lett.~B {\bf 606}, 123 (2005).

\bibitem{VMV} A.~Vogt, S.~Moch, J.~A.~M.~Vermaseren,  ``The three-loop splitting functions in QCD: The singlet case," Nucl.~Phys.~B {\bf 691}, 129 (2004).

\bibitem{Gelfand} I.~M.~Gel'fand and G.~E.~Shilov, ``Generalized Functions", Vol.~I, Sec.~3, Academic Press, New York (1964). We are dealing with the limiting case $\lambda\rightarrow -1$ of the construction in Sec.~3.8(2).

\bibitem{LDPutikka} The result of this construction appears directly without reference to generalized functions in an alternative probabilistic construction of the Altarelli-Parisi splitting functions in L.~Durand and W.~Putikka, Phys.~Rev.~D {\bf 36}, 2840 (1987). See  Eqs.~(19) in that paper. It would presumably also appear directly in a similar calculation of the coefficient functions above.

\bibitem{Diff_eq_G} M.~M.~Block, L.~Durand, and D.~W.~McKay, ``Analytic derivation of the leading-order gluon distribution function $G(x,Q^2)=xg(x,Q^2)$ from the proton structure function $F_2(x,Q^2)$," Phys.~Rev.~D {\bf 77}, 094003 (2008).

\bibitem{Analytic_solution_G} M.~M.~Block, L.~Durand, P.~Ha, and D.~W.~McKay, ``Analytic solution to the leading order coupled DGLAP evolution equations: A new perturbative QCD tool," Phys.~Rev.~D {\bf 83}, 054009 (2011).

\bibitem{Gribov} V.~N.~Gribov and L.~N.~Lipatov, ``Deep inelastic $ep$ scattering in perturbation theory,"  Sov.~J.~Nucl.~Phys. {\bf 15}, 438 (1972).

\bibitem{AltarelliParisi} G.~Altarelli and G.~Parisi, ``Asymptotic freedom in parton language," Nucl.~Phys.~B {\bf 126}, 298 (1977).

\bibitem{Dokshitzer} Y.~L.~Dokshitzer, ``Calculation of the structure functions for deep inelastic scattering and $e^+e^-$ annihilation in perturbation theory in quantum chromodynamics,"  Sov.~J.~Nucl.~Phys. {\bf 46},641 (1977).

\bibitem{H1_Coll_2011} F.~D. Aaron {\em et al.} (H1 Collaboration), ``Measurement of the inclusive $ep$ scattering cross section
at high inelasticity $y$ and of the structure function $F_L$,''  Eur.~Phys.~J. {\bf 71}, 1579 (2011).

\bibitem{H1_Coll_2014} V.~Andreev {\em et al.} (H1 Collaboration), ``Measurement of inclusive $ep$ cross sections at high $Q^2$ at
$\sqrt{s}= 225$ and 252 GeV and of the longitudinal proton
structure function $F_L$ at HERA,'' Eur.~Phys.~J. {\bf 74}, 2814 (2014).


\bibitem{Gradshteyn} I.~S.~Gradshteyn and I.~M.~Ryzhik, ``Table of Integrals, Series, and Products,'' Academic Press, New York (1965).



\end{thebibliography}
\end{document}